\documentclass[prx,twocolumn,preprintnumbers,amsmath,amssymb,showkeys]{revtex4}

\usepackage{graphicx}
\usepackage{bm}
\usepackage{color}

\begin{document}


\title{Modeling the relaxation processes of photoexcited solids: A short review}
\author{Shota Ono}
\email{shota_o@gifu-u.ac.jp}
\affiliation{Department of Electrical, Electronic and Computer Engineering, Gifu University, Gifu 501-1193, Japan}

\begin{abstract}
Ultrafast electron dynamics of solids after an absorption of femtosecond laser pulse is governed by electron-electron, electron-phonon, phonon-electron, and phonon-phonon collisions. It is of importance to construct a framework for interpreting experimental observations correctly. In this paper we review recent developments of modeling the relaxation dynamics of solids. We discuss the ultrafast relaxation in respect to the effective temperature dynamics and the excess energy dynamics. 
\end{abstract}

\keywords{electron-phonon collision; phonon-phonon collision; backward energy flow}


\maketitle

\section{Introduction}
Since the advent of femtosecond lasers, the ultrafast dynamics of solids has been extensively investigated with pump-probe spectroscopy: The first pump pulse excites the electrons into a far-from-equilibrium state, while the subsequent probe pulse examines the time-evolution of the photoexcited solid, where the electrons interact not only with each other but also with phonons. Given the observation of the ultrafast relaxation dynamics ranging from a femtosecond to nanosecond timescale, one important question is what physics we can obtain from a study of such relaxation dynamics. In this paper, we explore an answer to this question by considering the dynamics of the effective temperature (in Section \ref{sec:temperature}) and the excess electron and phonon energies (in Section \ref{sec:excess}). An effect of the hot phonons will be discussed in detail. We will also provide future perspectives in Sec. \ref{sec:summary}.

In this paper, we will focus on general features of the relaxation dynamics of excited solids. For a comprehensive review of recent advances on the hot electron dynamics of group IV and III-V semiconductors, see Ref.~\cite{sjakste}. 

\section{$N$-temperature dynamics}
\label{sec:temperature}
\subsection{$N=2$}
\begin{figure*}[ttt]
\center
\includegraphics[scale=0.45]{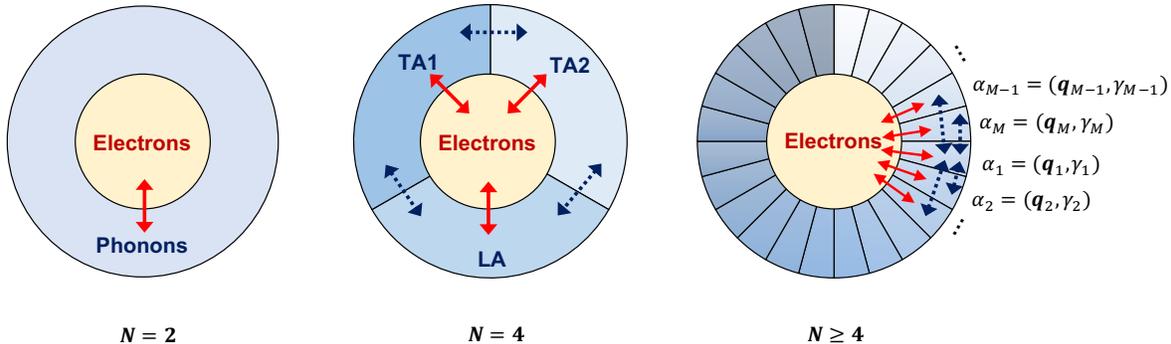}
\caption{\label{fig1} Schematic illustration of the $N$-temperature models for the cases of $N=2$, $N=4$, and $N\ge 4$. For $N=4$ there are three phonon subsets including the LA, TA1, and TA2 phonons, while for $N\ge 4$, there are $M$ phonon modes denoted as $\alpha_i =(\bm{q}_i, \gamma_i)$ with the wavevector $\bm{q}_i$ and the branch $\gamma_i$ ($i=1,\cdots, M$). Each region separated by curves is endowed with an effective temperature. The solid (red) and dashed (blue) arrows indicate the energy transfer through the e-ph and ph-ph collisions, respectively. }
\end{figure*}

Originally it was Allen \cite{allen} who had pointed out theoretically that the electron-phonon (e-ph) coupling constant can be obtained from an analysis of time-resolved experimental data. This idea is based on the two-temperature model (2TM) which can be derived from the Boltzmann equation \cite{ziman,smith} for the electron and phonon distribution functions, where the contribution from the diffusion is neglected because a very short timescale is considered. In the main assumption in deriving the 2TM, the electron-electron (e-e) and phonon-phonon (ph-ph) collision rates are large enough to keep electron and phonon quasiequilibrium states characterized by the electron temperature $T_e$ and phonon temperature $T_{ph}$, respectively, at any time. The excess electron energy is then transferred into the phonon system by the e-ph collisions until the system reaches a thermal equilibrium state (see the case of $N=2$ in Fig.~\ref{fig1}). Such dynamics is described by two-coupled differential equations with respect to $T_e$ and $T_{ph}$, known as 2TM,
\begin{eqnarray}
 C_e\frac{\partial T_e}{\partial t} = - C_{ph}\frac{\partial T_{ph}}{\partial t}
 = - G_{e-ph} (T_e - T_{ph}),
 \label{eq:2TM}
\end{eqnarray}
where $C_e$ and $C_{ph}$ are the specific heat of the electron and the phonon, respectively. The e-ph coupling factor $G_{e-ph}$ has been proved to be proportional to the e-ph coupling constant $\lambda \langle \omega^2 \rangle$ that is important in estimating the superconducting transition temperature of metals. Application of the 2TM \cite{allen} and/or an analytical solution to a simplified Boltzmann equation \cite{kabanov2008} to time-resolved experiments has been done in a variety of materials such as metals \cite{brorson}, superconductors \cite{gadermaier}, density-wave materials \cite{stojchevska}, and nanocarbons \cite{ono2014}, which enabled us to determine the e-ph coupling constant. The 2TM has been extended to study the ultrafast dynamics of the Dirac materials \cite{bistritzer,viljas,lundgren}. 

It is noteworthy that the coupling factor $G_{e-ph}$ in Eq.~(\ref{eq:2TM}) as a function of $T_e$ for various simple metals has been calculated based on the density-functional theory (DFT) \cite{lin}. The effect of the thermal excitation of $d$ band electrons could result in a significant change in the magnitude of $G_{e-ph}$, showing a limitation of the use of the free-electron gas model for the conditions of strong e-ph nonequilibrium.

\subsection{$N\ge 4$}
One may concern that the assumption behind the 2TM is questionable: The e-e and ph-ph collision rates are not large enough to establish the quasiequilibrium at any time. In fact, several theoretical and experimental works have shown the breakdown of the standard 2TM. In this subsection, we focus on recent advances in the temperature dynamics beyond 2TM.

First, we introduce a work done by Waldecker {\it et al}. \cite{waldecker}, suggesting that the presence of the nonequilibrium phonon distribution has a large impact on the energy flow between electrons and phonons. They have investigated the time-evolution of the atomic mean-squared displacement of aluminum experimentally. To interpret it, they started from the Boltzmann equation with the e-ph collision integral and derived four-temperature model (4TM) with the electron temperature, the longitudinal acoustic (LA) phonon temperature, and the two transverse acoustic (TA) phonon temperatures (the case of $N=4$ in Fig.~\ref{fig1}). The specific heat and the e-ph coupling factors $G_{e-ph}$ for each branch were taken from DFT calculations. The ph-ph coupling parameter was introduced to account for the energy transfer between three acoustic branches. One of the important implications in their work is that although both 2TM and 4TM can explain the time-resolved experimental data, the value of $G_{e-ph}$ obtained from the fit using 2TM is approximately by a factor of 2 smaller than that obtained from DFT calculations. In other words, the electron relaxation in the presence of quasiequilibrium phonon state is predicted to be faster than that in the presence of nonequilibrium phonon state. Sadasivam {\it et al}. \cite{sadasivam} and the author \cite{ono2018} have also shown that the use of 2TM would yield an underestimation of the e-ph coupling. This may be qualitatively explained by a constrained successive thermalization mechanism \cite{sadasivam}, where the number of phonons that can interact with electrons increases with the degree of the e-ph thermalization. 

The breakdown of the 2TM was a surprise because it happens even in the free-electron metal aluminum. Therefore, one of the current topics in the field of ultrafast dynamics has been to develop a model beyond the 2TM. Maldonado {\it et al}. have developed a multi-temperature model applicable to the ultrafast dynamics of alloys \cite{maldonado}. They have started from the Boltzmann equation considering the ph-ph as well as the e-ph collision integrals. Each phonon mode characterized by the wavevector and the branches ({\it i.e}., the optical as well as acoustic branches) is endowed with an effective phonon temperature (the case of $N\ge 4$ in Fig. 1). The multi-temperature model is derived by making an expansion of the distribution functions to second order in the temperature difference between the phonon mode and the electron. The e-ph and ph-ph coupling parameters were computed from DFT calculations. They applied the model to FePt and demonstrated that even after several tens of ps the phonons are not in thermal equilibrium since the effective phonon temperatures cover a range of hundreds of kelvins. This is due to the phonon mode dependent e-ph coupling and the mode dependent ph-ph scattering rates. The e-ph coupling strength is stronger for the optical mode than for the acoustic mode. The ph-ph scattering lifetime of the optical phonon mode ranges from 2 to 10 ps, while that of the acoustic mode is larger than 10 ps. Such parameter differences could yield the slow relaxation dynamics. This study implies that the higher the degree of freedom in the system, the slower the relaxation dynamics. 

\section{Excess energy dynamics}
\label{sec:excess}
In the preceding section, we have seen that the concept of the time-dependent temperature is useful to interpret the ultrafast electron and phonon dynamics, while a significant modification of the 2TM is necessary for a quantitative understanding of them. Another way toward a unified understanding of nonequilibrium dynamics is to study the time-evolution of the excess electron energy measured from the energy stored before an excitation.

\begin{figure}[ttt]
\center
\includegraphics[scale=0.45]{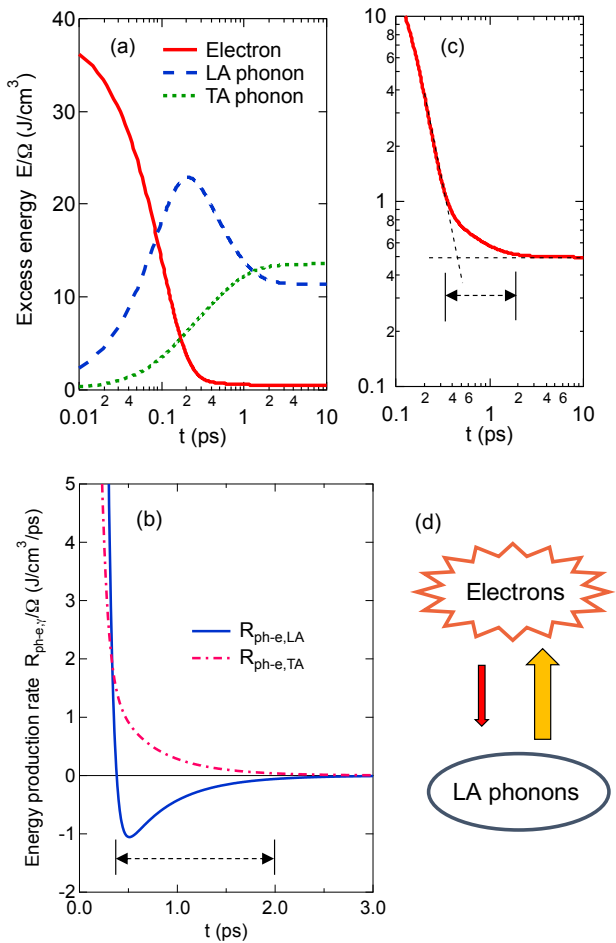}
\caption{\label{fig2} (a) The time-evolution of the excess energies of the electron (solid), LA phonon (dashed), and TA phonon (dotted) in an electronically excited simple metal. (b) The energy production rate of the LA (solid) and TA (dot-dashed) phonon energy as a function of $t$. (c) The magnified view of the curve of the excess electron energy versus $t$. The dashed arrows in (b) and (c) indicate the same interval of $t=0.3$-$2$ ps. (d) Schematic illustration of the backward energy flow. The width of arrows indicates the amount of the transferred energy. Figures (a)-(c) extracted and edited from Ref.~\cite{ono2018}. }
\end{figure}

\begin{figure}[ttt]
\center
\includegraphics[scale=0.45]{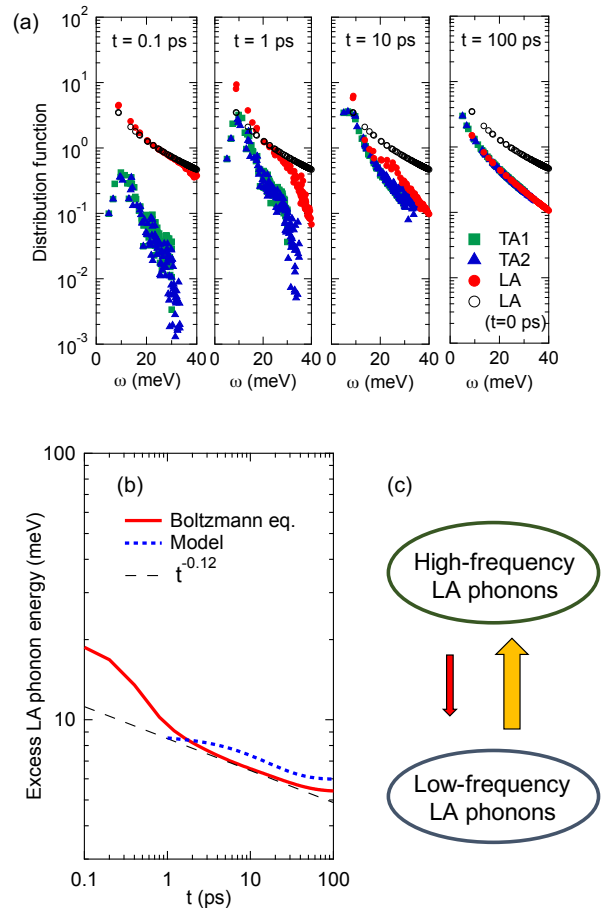}
\caption{\label{fig3} (a) The nonequilibrium phonon distribution for TA1 (square), TA2 (triangle), and LA (filled circle) phonons. The distribution of the LA phonon at $t=0$ ps is also shown. (b) The time-evolution of the LA phonon energy obtained from a solution to the Boltzmann equation (solid) and simplified differential equation (dotted). The thin curve (dashed) is a power law decay $t^{-p}$ with $p=0.12$. (c) Schematic illustration of the backward energy flow for phonons. Figures (a)-(b) extracted and edited from Ref.~\cite{ono2017}. }
\end{figure}

\subsection{Final stage relaxation dynamics}
Recently, the author has reported a series of studies \cite{ono2018,ono2017} focusing on the final stage of the relaxation in simple metals and found an anomalous relaxation behavior. In Ref.~\cite{ono2018}, we solved the Boltzmann equation with consideration of the e-e, e-ph, and ph-ph collision integrals and obtained the time-evolution of the electron and phonon distribution functions, from which the excess electron and phonon energies were calculated  (see Fig.~\ref{fig2}(a)). In the initial stage of the relaxation for aluminum, the excess electron and phonon energy decreases and increases with time, respectively. This shows the electron energy transfer into the LA and TA phonons through the e-ph collisions. Simultaneously, the excess LA phonon energy is transferred into TA phonons through three-phonon collision events. Such an interplay between the e-ph and ph-ph collisions yields an overshoot of the LA phonon energy at a specific time of 0.2 ps, after which the system gradually reaches a thermal equilibrium state. Interestingly, some parts of the LA phonon energy are inversely transferred to the electron system in the final stage of the relaxation. This is shown as the negative value of the energy production rate of the LA phonon through the phonon-electron (ph-e) collisions (Fig.~\ref{fig2}(b)), where the energy production rate is defined as the time-derivative of the excess phonon energy due to the ph-e collisions. During the inverse energy flow, the time-evolution of the excess electron energy shows an anomalously slow dynamics (Fig.~\ref{fig2}(c)). The strong correlation between the backward energy flow (Fig.~\ref{fig2}(d)) and the slow electron dynamics is one of the main findings in Ref.~\cite{ono2018}.

The slow relaxation together with the backward energy flow has also been observed in more simplified phononic systems \cite{ono2017}. We studied nonequilibrium phonon dynamics in a monatomic face-centered cubic lattice, where the three-phonon scattering rates were modeled within continuum elasticity theory \cite{landau}. We set the initial condition that the LA phonon temperature ($k_{\rm B}T=35$ meV with $k_{\rm B}$ the Boltzmann constant) is higher than the TA phonon temperatures ($k_{\rm B}T=1$ meV) and computed the time-evolution of LA and TA phonon distribution functions through numerical integration of the Boltzmann equation, as shown in Fig.~\ref{fig3}(a). The energy is transferred from the LA to TA1 and TA2 phonons, yielding a thermal equilibrium with $k_{\rm B}T=17$ meV at $t=$100 ps. Similar to the results in Ref.~\cite{ono2018}, the excess LA phonon energy decreases quite slowly in the final stage of the relaxation, while in this study the energy shows a power law behavior $\propto t^{-p}$ with the exponent of $p=$0.12 (Fig.~\ref{fig3}(b)). This relaxation dynamics occurs due to a backward energy flow from the low-frequency (around the $\Gamma$ point in the Brillouin zone) to the high-frequency phonons (Fig.~\ref{fig3}(c)), which happens subsequent to an overheating of the low-frequency phonons due to the high-frequency phonon decay in the initial relaxation. The main physics behind the phonon energy transfer has been successfully described by a simple differential equation for an effective temperature of the low-frequency phonons. For mathematical details, see Ref.~\cite{ono2017}. 

\subsection{trARPES}
The time- and angle-resolved photoemission spectroscopy (trARPES) has allowed us to study the time-evolution of the electron distribution function of photoexcited solids and extract the excess energy. From an experimental point of view, it is also of fundamental importance how the nonequilibrium electron distribution evolves into the Fermi-Dirac (FD) distribution. In earlier studies using time-resolved photoemission spectroscopy, the FD formation dynamics has been investigated for simple metals \cite{fann,lisowski} and graphite \cite{ishida2011}. The breakdown of 2TM has been confirmed by direct observations of the distribution function.

The recent studies using trARPES on SrMnBi$_2$ \cite{ishida2016} and black phosphorus \cite{ishida2018} have shown that the excess electron energy shows a slow dynamics, that is, a power law decay just before reaching a thermal equilibrium. For the latter, an exciton effect on the electron relaxation has been suggested to explain the power law decay. Nevertheless, we consider that the final relaxation observed experimentally may be understood by a concept of the backward energy flow \cite{ono2018,ono2017}, for example, between the bound states and the free carriers, in a unified manner.

Quite recently, Rohde {\it et al}. has investigated the excess energy dynamics for graphite by using trARPES \cite{rohde}. They determined an effective electron temperature $T_e$ from a fit of a FD distribution to the experimental data and compute the excess energy $\Delta E$ stored in the conduction band at each time. The plot of $\Delta E$ versus $T_e$ has revealed that nonequilibrium electron dynamics can be dissected into substages: the photon absorption, the momentum redistribution of the excited electrons by the e-e collisions, and the cooling of the electrons due to the e-ph collisions. Such an analysis would be helpful to ease the complexity of nonequilibrium dynamics, when momentum and energy relaxation occurs at different timescales.  

Using a nonequilibrium Green's function formalism, the relaxation dynamics of photoexcited e-ph systems, where the Holstein Hamiltonian is used to describe the interaction between the electrons and phonons, have been studied to understand the trARPES data \cite{sentef,abdurazakov}. Recent study in Ref.~\cite{abdurazakov} has shown that the excited phonon populations cause the quasiparticle decay rates to decrease. This can also be understood as the energy transfer from the phonon to electron systems through the phonon absorption.

\section{Summary and future perspectives}
\label{sec:summary}
We have reviewed recent studies of nonequilibrium dynamics of electrons and phonons in photoexcited solids and tried to answer what the main physics is in the relaxation dynamics. The effective temperature (in Section II) and the excess energy (in Section III) play an important role for understanding the ultrafast dynamics in a variety of materials. These are summarized as follows:
\begin{enumerate}
\item The use of the well-known 2TM would lead to an underestimation of the e-ph coupling constant. 
\item The observation of the slow decay of the excess electron energy in the final relaxation indicates a backward energy flow from X to Y systems, where X and Y are, for example, the phonon and the electron, respectively. 
\end{enumerate}
It is obvious that the nonequilibrium treatment for phonons is required in reaching these conclusions. More generally, the dynamics involving many degrees of freedom, not two, must be examined to develop a theory for nonequilibrium phenomena of solids. Work along this line is in progress.

There have been much works in respect to nonequilibrium dynamics of quasiparticles and other elementary excitations on superconductors and semiconductors. The nonequilibrium dynamics of them is also interesting, partly because other phenomenological models different from the 2TM have been proposed for interpreting the ultrafast dynamics in each system and partly because a deep understanding of the carrier dynamics in particularly semiconductors is essential in operating electronic and optoelectronic devices. For example, the Rothwarf-Taylor (RT) model can describe the low-energy dynamics of superconducting quasiparticle \cite{RT}. Such a quasiparticle dynamics forming a Cooper pair is strongly influenced by the phonon emission and absorption processes. The relaxation dynamics is thus governed by the anharmonic decay of high-frequency phonons into low-frequency phonons with the energy enough not to excite the quasiparticle above the superconducting energy gap \cite{kabanov1999,ono2012}. As well as the low-energy dynamics, modeling the high-energy quasiparticle dynamics, leading to an energy gap suppression \cite{madan}, would be desired to understand the nonthermal melting of superconductors. 

We have considered the time-evolution of the distribution function represented in $k$-space and driven by weak excitations. Modeling ultrafast laser heating in real space and under an intense photon flux is important to study the laser ablation of solids \cite{rethfeld}. An atomistic approach such as molecular dynamics simulation, as well as the Boltzmann equation approach, would play a crucial role toward a unified understanding of nonequilibrium dynamics of condensed matter systems.

\begin{acknowledgments}
The author thanks M. Yoshiya for giving me a chance to write this review and Y. Ishida, T. Otobe, R. Kobayashi, and S. Kato for stimulating discussions.
\end{acknowledgments}

\end{document}